\documentclass[a4paper,10pt]{article}
\usepackage[utf8]{inputenc}
\usepackage{amsmath}
\usepackage{amsfonts}
\usepackage{amssymb}
\usepackage{graphicx}
\usepackage{epstopdf}

\title{
Two photon amplitude of partially coherent partially entangled electromagnetic fields
}
\author{Miguel Angel Olvera, Sonja Franke-Arnold}

\begin{document}

\maketitle


\begin{abstract}
\noindent
The development of efficient protocols for pure and mixed states preparation is challenging task. Most of the theory of quantum information applications has been developed for fully coherent or completely incoherent light. However, in many situations of interest partially coherent light has been proved to be a more robust model of radiation.	
In this paper the underpinning theory of two photon amplitude functions for down-converted fields with partially coherent pump beams is investigated. By using the generalised Siegert relations and the coherent mode representation of the cross spectral density matrix the two photon amplitude is fully characterised for partially coherent beams. A number of correlation properties from modern coherence theory are demonstrated to be preserved under parametric down-conversion. 
Based on the generalised Siegert relations and the Cauchy-Schwarz inequality a measure of entanglement for the two photon amplitude is proposed. Upper bounds for this measure are found in terms of the \emph{golden ratio} for maximally entangled states. Two inequalities are derived for the two photon amplitude for which there exist a transition zone from super-Poisson statistics to sub-Poisson statistics for down-converted partially coherent fields.
\end{abstract}

\section{Introduction}
\noindent
The two photon amplitude (TPA) plays a role of paramount importance in the concept of quantum entanglement which has been largely recognized as a fundamental property of the non linear phenomena known as parametric down conversion (PDC) \cite{PhysRevA.31.2409}.
The widespread use of bi-photon generation techniques and the improvement of coincidence detection of photons has led to the understanding of the entanglement of bipartite quantum systems as well as to the implementation of efficient quantum information protocols.   	
\noindent
However in the vast majority of investigations, the pump field is assumed to be coherent and the strategy consists in tailoring the TPA to investigate changes in the spectral properties of bipartite systems that can generate high dimensional entanglement or lead to new entanglement schemes \cite{PhysRevA.57.3123,PhysRevA.54.5349,PhysRevA.67.052313,PhysRevLett.92.127903,PhysRevA.65.033823}. The aim of the present paper is to include the correlation properties of the pump field in order to investigate correlation-induced entanglement effects. 

TPA shape manipulation has been investigated by two main approaches, by changing the pump spectrum and by  adjusting the phase matching conditions.
The angular spectrum of the pump laser, preserved it the two photon state, allows taking control of the transverse  correlations by transferring the image from the pump to the down-converted field cite{PhysRevA.57.3123}.  
Focusing a laser pump field has led to the evaluation of the correlation function in an imaging system with 
two photon geometric optics effects by passing the laser beam through a focusing lens before the down-conversion crystal \cite{PhysRevA.53.2804}.
\noindent
Down-converted photons have been used to investigate the effect of pump spectrum and the various parameters of entangled-photon imaging systems on the imaging resolution \cite{Abouraddy:02}.
Entanglement enhancement has been investigated on the basis of a judicious selection of eigenmodes in angular momentum entanglement \cite{PhysRevA.65.033823}. 
The characterization of entanglement by determining an appropriate set of bi-orthogonal mode pairs has been performed by using the Schmidt decomposition of the TPA \cite{PhysRevLett.84.5304}. 
This approach has been used for engineering quantum correlations by controlling spectral amplitudes \cite{walmsley03}.
The eigenvalues and eigenmodes of the density matrix have been used to approximate the TPA by a bi-Gaussian function  and for calculating the Schmidt number \cite{fedorov09}.

The two photon state can be controlled by using transversely patterned quasi-phase matching gratings to tailor the spatial mode function that describes the state of photon pairs in PDC \cite{Torres:04}. It has been shown that type-II down-converted photons with broadband pump exhibit a continuous superposition of two-photon states in which the probability amplitude for each state depends on the pump bandwidth whose envelope function determines the range of pump energies and the phase-matching function determines the energy spectrum \cite{PhysRevA.56.1627}.

In all the approaches mentioned above, the common characteristic is the use of coherent sources from down-converted fields are obtained. An important number of potential sources, namely, partially coherent sources have not been included and only a few studies have put forward the connection between entanglement and partial coherence. The mathematical theory of duality between partial coherence and entanglement comprises the description of propagation, diffraction and interference of bi-photon fields in optical linear systems \cite{PhysRevA.62.043816}.
Coherence properties of down-converted fields were investigated in terms of the physical properties of a coherent pump beam \cite{PhysRevA.53.4360}. In this connection, recently it was proved that down-converted photons retain the coherence properties of the pump field and the maximum entanglement is bounded by the degree of spatial coherence \cite{PhysRevA.81.013828}. With the development of modern coherence theory, in the last decade a number of results and techniques have been established that brings deep insight into the concepts of coherence and polarisation expressed through the cross spectral density matrix \cite{wolf2007introduction,vct,Gori:07,Gori:09,Tervo:03,Ostrovsky:09,Ostrovsky:09,Ostrovsky:10,1464-4258-7-5-004,PhysRevA.89.013801}. 

In the present paper theoretical implications of correlation-induced entanglement effects, i.e., tailoring the TPA photon amplitude with partially coherent fields on bipartite partially entangled systems, are investigated. The paper is organized as follows. In section 2 it is demonstrated that the coherence properties of partially coherent pump fields are preserved in the two photon entangled fields. In section 3 the cross spectral density operator is introduced. The expected value of this operator is related to the two photon amplitude trough the kernel of the one photon amplitude. In section 4 an extension to the measure of entangled states proposed in \cite{PhysRevA.64.050101} is established for the TPA of partially coherent fields that satisfies the Siegert relation. Under this formalism it is found that maximally entangled states have an upper bound given by the golden ratio. An important consequence is that entanglement beyond the golden ratio implies the violation of the Cauchy-Schwarz inequalities for maximally entangled states and a transition from super-Poisson statistics to sub-Poisson is involved. Finally, in section 5 conclusions are presented.

\section{Coupling of partially coherent down-converted fields}

The TPA contains all the information of the down converted fields field at the moment of birth inside the non-linear material. The correlation time of the pump field was believed to play no role in the two photon second order correlations, fundamentally because in Mandel's original derivation the bandwidth was taken to be zero and in part because the concepts of partial coherence in the space frequency domain was not fully developed. Note that Wolf's theory of partial coherence in the space frequency domain was published a year later than Mandel's theory, in 1986 \cite{PhysRevA.31.2409,Wolf:86}.  
In the vast majority of investigations concerned with two photon systems, the pump beam is assumed to be coherent. If the pump is partially coherent the two-photon field operator depends on the interaction time, the representation of the classical field and the non linear properties of the crystal. The expected value of the field operator depends on both the expected value of the quantum field operator and the cross spectral density matrix.
Partially coherent fields pertain to a large class of electromagnetic sources that obey super-Poisson statistics although not fully unbiased, they are often referred to thermal fields. 

Among the large class of partially coherent sources, here we are interested in partially coherent fields for which the fourth order correlation function $\Gamma^{(2)}(\mathbf{r}_{1},\mathbf{r}_{2})
$ can be expressed through the second order correlation function, i. e. those for which the Siegert relation holds \cite{PhysRevLett.94.223601}

\begin{eqnarray}\label{1}
\Gamma^{(2)}(\mathbf{r}_{1},\mathbf{r}_{2})
=
\Gamma^{(1)}(\mathbf{r}_{1},\mathbf{r}_{1})
\Gamma^{(1)}(\mathbf{r}_{2},\mathbf{r}_{2})
+
\vert
\Gamma^{(1)}(\mathbf{r}_{1},\mathbf{r}_{1})
\vert ^{2}.
\end{eqnarray}

With this relations the fourth order correlation functions are characterised in terms of second order correlation functions. 

\noindent
In the following the down-conversion of photons from the partially coherent pump field will be analysed following Mandel's original derivation \cite{PhysRevA.31.2409} within the time where the interaction takes place. The analysis is valid for non-degenerate type I and type II SPDC. 
Consider an electromagnetic field $V_{l}(\mathbf{r},t)$ entering a dielectric non linear crystal with bilinear susceptibility $\chi$, the Hamiltonian of the system is
	
\begin{equation}
H=
\frac{1}{2}
\int_{D}\chi_{ijl}
\hat{U}_{i}(\mathbf{r},t)\hat{U}_{j}(\mathbf{r},t)V_{l}(\mathbf{r}_{0},t)
\mathrm{d}\mathbf{r},
\end{equation} 	
\noindent
where $D$ is the spatial domain where the interaction of light and the crystal takes place and $V_{l}(\mathbf{r}_{0},t)=V_{l}e^{[i(\mathbf{k}_{0}\cdot\mathbf{r}_{0}-\omega(\mathbf{k}_{0})t)]}$ is a member of the classical pump field ensemble of realizations with average frequency $\omega_{0}$. The quantized field is given by

\begin{equation}
\hat{U}(\mathbf{r},t)
=
\frac{1}{L^{3/2}}
\int_{D} 
\varepsilon_{\mathbf{k}s}
\hat{a}^{\dagger}_{\mathbf{k}}(\mathbf{k},t)
e^{i(\mathbf{k}\cdot \mathbf{r})}
\mathrm{d}\mathbf{k},
\end{equation} 	
\noindent
where $\hat{a}^{\dagger}(\mathbf{k},t)$ is the field creation operator of the mode $\mathbf{k}$, $\varepsilon_{\mathbf{k}s}$ is a unit polarisation vector depending on the wavevector $\mathbf{k}$, $s=1,2$ is the polarization index and $\mathbf{r}$ spans the transverse plane perpendicular to the propagation axis. The Hamiltonian is then given by

\begin{align}
\hat{H}
=
\int_{D}
\int	\int
\chi_{ijl} 
\hat{a}_{\mathbf{k}s}^{\dagger}
\hat{a}_{\mathbf{k}'s'}^{\dagger}
V_{l}(\mathbf{r}_{0},t)  
\varepsilon_{\mathbf{ks}_{i}}^{*}
\varepsilon_{\mathbf{k's'}_{j}}^{*} \notag \\
e^{i
[(\mathbf{k}_{0}-\mathbf{k}-\mathbf{k}')\mathbf{r}
-\omega(\mathbf{k}_{0})t]}
\mathrm{d}\mathbf{r}
\mathrm{d}\mathbf{k} 
\mathrm{d}\mathbf{k}'.
\end{align}
\noindent
By defining $\hat{A}=1/L^{3/2}\hat{a}e^{i\omega(\mathbf{k})t}$ the Heisenberg equations of motion can be written as follows 

\begin{align}
\dot{\hat{A}}
=&
\frac{1}{i\hbar L^{3}}
\int_{D}
\int
\int
\chi_{ijl} 
\epsilon_{\mathbf{k}s,\mathbf{k}'s'}
\hat{A}_{\mathbf{k}'s'}^{\dagger} 
V_{l}(\mathbf{r}_{0},t) \\ \notag
&
e^{i[(\mathbf{k}_{0}-\mathbf{k}-\mathbf{k}')\cdot \mathbf{r}]}	
e^{-i[(\omega(\mathbf{k}_{0})-\omega (\mathbf{k})-\omega(\mathbf{k}'))t]}
\mathrm{d}\mathbf{r}
\mathrm{d}\mathbf{k}
\mathrm{d}\mathbf{k}',
\end{align}
\noindent
where $\epsilon_{\mathbf{k}si,\mathbf{k}'s'j}$ represents the unit polarisation vector. In order to determine the field mode the Heisenberg equations of motion must be integrated over the interaction time $\Delta t$. If the pump field is considered coherent with a coherence time of the order of microseconds and since the interaction time is typically a fraction of a nanosecond the time integral is straightforward and yields the familiar sinc function, whose argument contains the phase matching conditions, of the frequency and the interaction time.
\noindent
In the case of partially coherent sources the situation is different. The values of temporal coherence ranges from a few to hundreds of femtoseconds for partially coherent sources and fractions of nanoseconds for multimode lasers. With the development of the modern coherence theory \cite{Gbur2010285}, an important number of sources with prescribed properties of coherence has been put forward so that the range of temporal coherences available under the time integration is non negligible. Hence the integration over time is expressed as follows

\begin{align}
\hat{A}
=&
\frac{1}{i\hbar L^{3}}
\int_{\delta t}
\int_{D}
\int
\int
\chi_{ijl} 
\hat{A}_{\mathbf{k}s}^{\dagger}	
T_{ij}(\mathbf{k}s,\mathbf{k}'s') 
V_{l}(\mathbf{r}_{0},t') \\ \notag
&
e^{i[(\mathbf{k}_{0}-\mathbf{k}-\mathbf{k}')\cdot\mathbf{r}]}
e^{-i[(\omega(\mathbf{k}_{0})-\omega (\mathbf{k})-\omega(\mathbf{k}'))t')]}
\mathrm{d}\mathbf{r}
\mathrm{d}\mathbf{k}
\mathrm{d}\mathbf{k}'
\mathrm{d}t'.
\end{align}
\noindent
It is convenient to express this integral as follows

\begin{align}
\hat{A}
=&
\frac{1}{i\hbar L^{3}}
\int
\int
\Big[
\int_{\delta t}
V_{l}(\mathbf{r},t')
e^{-i[\omega(\mathbf{k}_{0})-\omega(\mathbf{k})-\omega(\mathbf{k'})]t'}
\mathrm{d}t'
\Big]
\Big[
\int_{D}
\hat{A}_{\mathbf{k}s}^{\dagger}
e^{-i[\mathbf{k}_{0}-\mathbf{k}-\mathbf{k}']\mathbf{r}}
\mathrm{d}\mathbf{r}
\Big] \notag
\\
&	
\chi_{ijl}
T_{ij}(\mathbf{k}s,\mathbf{k}'s') 
\mathrm{d}\mathbf{k}
\mathrm{d}\mathbf{k}',
\end{align}
\noindent
where $\delta t$ corresponds the interaction time. If the pump field is partially coherent the contributions to the response operator of down-converted light at point $\mathbf{r}$ and time $t$ resulting from the creation operator will be affected by the correlations of the input beam within the non-linear material. To show how this happen, consider the operator 

\begin{align}
\mathbf{U}(\mathbf{r},t)
=
\mathbf{U}_{free}(\mathbf{r},t)
+
\int
\hat{\mathcal{A}}_{\mathbf{k}s}
\tilde{V}(\mathbf{r},\omega)
F(\mathbf{k},s,\mathbf{r},t)
\mathrm{d}\mathbf{k},
\end{align}
\noindent
where $\mathbf{U}_{free}$ is the field operator in the absence of any interaction, $\tilde{V}(\mathbf{r},\omega)$ is the Fourier spectrum of the input field within the interaction interval $\delta t$ 

\begin{eqnarray}
\tilde{V}(\mathbf{r},\omega)
=
\int_{t_{0}}^{t_{0}+\delta t}
V_{l}(\mathbf{r},t)
e^{-i[\omega(\mathbf{k}_{0})-\omega(\mathbf{k})-\omega(\mathbf{k'})]t}
\mathrm{d}t,
\end{eqnarray},
\noindent
\begin{eqnarray}
\hat{\mathcal{A}}_{\mathbf{k}s}
=
\int_{}^{}
\hat{A}^{\dagger}_{\mathbf{k}s}
e^{-i[\mathbf{k}_{0}-\mathbf{k}-\mathbf{k}']\mathbf{r}}
\mathrm{d}\mathbf{r},
\end{eqnarray}
\noindent
and $F(\mathbf{k},s,\mathbf{r},t)$ is given by

\begin{align}
F(\mathbf{k},s,\mathbf{r},t)
=&
\frac{1}{i\hbar L^{3}}
\int
\int
\chi_{ijl} 
T_{ij}(\mathbf{k}s,\mathbf{k}'s') 
\mathrm{d}\mathbf{k}
\mathrm{d}\mathbf{k}'.
\end{align}
\noindent
It was pointed out in the previous section that, by virtue of the Siegert relation $\Gamma^{(2)}(\mathbf{r}_{1},\mathbf{r}_{2})
=
\Gamma^{(1)}(\mathbf{r}_{1},\mathbf{r}_{1})
\Gamma^{(1)}(\mathbf{r}_{2},\mathbf{r}_{2})
+
\vert
\Gamma^{(1)}(\mathbf{r}_{1},\mathbf{r}_{1})
\vert ^{2}$, the fourth order correlation function of partially coherent beams is fully described by the second order correlation function. From the present analysis is clear that the averaged measure of light intensity, at points $(\mathbf{r}_{1},t)$ and $(\mathbf{r}_{2},t)$ for $t > \delta t$, is given by 
$\Gamma^{(2)}(\mathbf{r}_{1},\mathbf{r}_{2})=\langle \mathbf{U}^{\dagger}(\mathbf{r}_{1},t)
\mathbf{U}^{\dagger}(\mathbf{r}_{2},t)
\mathbf{U}(\mathbf{r}_{2},t)
\mathbf{U}(\mathbf{r}_{1},t)
\rangle$, therefore it is fully characterised by the second order correlation function given by

\begin{eqnarray}\label{g21}
\Gamma^{(1)}(\mathbf{r}_{1},\mathbf{r}_{2};\omega)
=
\int_{\mathbf{k}}
\int_{\mathbf{k}'}
\langle 
\hat{\mathcal{A}}_{\mathbf{k}s}^{\dagger}
\hat{\mathcal{A}}_{\mathbf{k}'s'}
\rangle
\langle 
\tilde{V}^{*}(\mathbf{r}_{1},\omega)
\tilde{V}(\mathbf{r}_{2},\omega)
\rangle \notag \\
F^{*}(\mathbf{k},s,\mathbf{k'},s',\mathbf{r}_{1},t)
F(\mathbf{k},s,\mathbf{k'},s',\mathbf{r}_{2},t)
&
\mathrm{d}\mathbf{k}
\mathrm{d}\mathbf{k}'.
\end{eqnarray}
\noindent
The first term in the integrand governs the quantum system of down converted photons, it depends on the vacuum expectation value of the boson operators during the interaction time and the phase matching conditions. The second term, recognised as the cross spectral density function of an electromagnetic field, is the correlation function that governs the spatial coherence properties of the classical pump field at frequency $\omega$. Therefore the expected value of the two photon amplitude contains the information about both the down converted photons and the classical correlation properties of the pump field.
\noindent
This result reveals the dual world to which down-converted photons belongs, when the pump field is partially coherent the contribution to the two photon amplitude is much richer than the previous case. We shall see in the next section that in order to quantify the effect of field correlations in down-converted photons the cross spectral density matrix can be conveniently expressed as a density operator allowing the evaluation of the second order correlation function.

\section{The cross spectral density operator}

It is well known that the second order correlation function satisfies the Wolf equations and the Helmholtz equation. The Wolf equations have a quantum counterpart whose solution are the TPA; as a consequence the fourth order correlation function also satisfies the Helmholtz equation \cite{PhysRevLett.94.223601}. This remarkable result constitutes a further demonstration of the duality between two photon systems and partially coherent fields. These equations posses identical mathematical structure in both position and Fourier space, besides both share the hallmark of quantum-optical systems namely, linearity. The solutions are therefore connected and describe  mathematically equivalent physical phenomena. The second order coherence function represents the oscillatory behaviour of partially coherent light in propagation and diffraction. The two photon amplitude represents the probability amplitude of oscillation modes in the system  of down-converted photons. Therefore the two photon Wolf equations imply that the TPA admits a modal representation in terms of the eigenvalues and eigenvectors of the second order correlation function. 

Despite of this compatibility further clarification is necessary. The second order correlation function and the cross spectral density function are correlation functions with analogous mathematical definition, however do not describe the same optical phenomena. The Young interference experiment reveals the field source correlations at the plane of interference. The mathematical description of the interference pattern involves two types of coordinates, i.e., the intensity, in terms of coordinates from the plane of interference, is weighted by the degree of coherence in terms of coordinates from the source plane. This subtle distinction reveals the nature of the coherence functions as source correlation functions. The degree of coherence does not provide any information of correlations that may occur at the interference plane, where the light is detected. The second order correlation function describes correlations of arrivals, i.e., detection correlations and the same holds for the fourth order correlation function. In this regard do not contain information about correlations at the source plane because they are in terms of photon counting statistics. For the sake of comparison it is possible to calculate the degree of coherence at the interference plane, the result is that in general is different from the source. The same holds in the other case, intensity correlations can be measured at the source plane but in general these will not be equal to the measurements at the detection plane. 

In the following we proceed to elucidate the connection between the second order correlation function and the cross spectral density matrix in the context of spontaneous parametric down conversion for partially coherent light. One of the most important results in vector coherence theory states that when the cross-spectral density (CSD) matrix satisfies square integrability:

\begin{eqnarray}\label{w1}
\int\int _{D}
\mathrm{Tr}
\Big [
\mathbf{W}^{\dagger}(\mathbf{r}_{1},\mathbf{r}_{2};\omega)	
\mathbf{W}(\mathbf{r}_{1},\mathbf{r}_{2};\omega)
\Big]
\mathrm{d}\mathbf{r}_{1}
\mathrm{d}\mathbf{r}_{2}
<
\infty ,
\end{eqnarray}
\noindent
hermitian symmetry:

\begin{eqnarray}\label{w2}
\mathbf{W}^{\dagger}(\mathbf{r}_{1},\mathbf{r}_{2};\omega)	
=
\mathbf{W}(\mathbf{r}_{2},\mathbf{r}_{1};\omega),
\end{eqnarray}

\noindent
and is non-negative definite:

\begin{eqnarray}\label{w3}
\int\int _{D}
\mathbf{f}(\mathbf{x}_{1})
\mathbf{W}(\mathbf{r}_{1},\mathbf{r}_{2};\omega)	
\mathbf{f}(\mathbf{x}_{2})
\mathrm{d}\mathbf{r}_{1}
\mathrm{d}\mathbf{r}_{2}
\geq
0,
\end{eqnarray}
\noindent
where $\mathbf{f}(\mathbf{r})$ is an arbitrary row matrix, hence the CSD becomes a Hilbert-Schmidt operator. This is an important result because the CSD functions can be expanded in Mercer's series \cite{Gori:03,Tervo:04}

\begin{eqnarray}\label{ms}
\mathbf{W}(\mathbf{r}_{1},\mathbf{r}_{2};\omega)	
=
\sum_{n}
\Lambda_{n}
\mathbf{\Phi}_{n}^{\dagger}(\mathbf{r}_{1})
\otimes
\mathbf{\Phi}_{n}(\mathbf{r}_{2}),
\end{eqnarray}
\noindent
where coefficients $\Lambda_{n}$ are the eigenvalues and $\mathbf{\Phi}_{n}(\mathbf{r})=[\varphi_{n,i}(\mathbf{r})]$ are the eigenvectors of the Fredholm integral equation given by

\begin{eqnarray}\label{fi}
\int_{D}
\mathbf{\Phi}_{n}(\mathbf{r}_{1})
\mathbf{W}(\mathbf{r}_{1},\mathbf{r}_{2};\omega)	
\mathrm{d}\mathbf{r}_{1}
\mathrm{d}\mathbf{r}_{2}
=
\Lambda_{n}
\mathbf{\Phi}_{n}(\mathbf{r}_{2}).
\end{eqnarray}
\noindent
Because of relations \eqref{w1}, \eqref{w2} and \eqref{w3} the eigenvalues $\Lambda_{n}$ are real and non-negative, and the eigenvectors $\mathbf{\Phi}_{n}(\mathbf{r})$ are orthonormal. Therefore, decomposition \eqref{ms} represents the cross spectral density matrix of any partially coherent and partially polarised electromagnetic field as a superposition of completely coherent completely polarised modes. The vector integral equation \eqref{fi} represents a set of two coupled scalar integral equations whose kernel $\mathbf{W}(\mathbf{r}_{1},\mathbf{r}_{2};\omega)$ is for the general case unknown. However considerable simplification is obtained when the kernel is diagonalised, in such a case the integral equations uncouple and can be treated independently.

Solving the integral equation \eqref{fi},for any particular case is not an easy task. For those cases in which this can be achieved, a fundamental property of Mercer's series is that they are absolutely and uniformly convergent. The eigenvalue distribution typically can involve a large number of elements, this fact makes possible to express the $m$th-eigenvalue for the $i$th-component of matrix \eqref{ms} in terms of power series. By the central limit theorem, the expansion of the $m$th-eigenvalue is of the form 

\begin{eqnarray}\label{la}
\Lambda _{m,i}(\lambda)
=
\sum_{l=0}^{m}
(-1)^{l}
\dfrac{\lambda_{i}^{2l}}{l!}.
\end{eqnarray}
\noindent
The parameter $\lambda_{i}$, that satisfies $0\leq \lambda_{i} \leq 1$, depends on the ratio between the effective width of the source $\sigma_{s}$ and the correlation length  ${\sigma_{c}}$ in a functional form determined by the particular type of source. 
\noindent
The power series expansion represents the contribution, per number of photons, of the $m$th-mode to the coherence of the field. The second order correlation function expressed in terms of the density operator $\hat{\rho}$ is given by

\begin{eqnarray}
\langle
\tilde{V}_{i}^{*}(\mathbf{r}_{1};\omega) \tilde{V}_{j}(\mathbf{r}_{2};\omega)
\rangle
=
\sum_{q}
\langle \phi_{q} \vert
\hat{\rho} \tilde{V}_{i}^{*}(\mathbf{r}_{1};\omega) \tilde{V}_{j}(\mathbf{r}_{2};\omega)
\vert \phi_{q} \rangle ,
\end{eqnarray}
\noindent
where

\begin{eqnarray}
\hat{\rho}
=
\sum_{n=0}^{\infty}
P_{n}(\alpha)
 \vert \alpha \rangle
\langle \alpha \vert .
\end{eqnarray}
\noindent
Therefore the modal expansion determines uniquely the contribution of modes per number of photons to the coherence of the field. The eigenvalue expansion in the coherent mode representation is related to the coherent state representation as follows

\begin{eqnarray}\label{lfi}
\sum_{q}
\langle \phi_{q} \vert
\rho \tilde{V}^{*}_{i}(\mathbf{r}_{1};\omega) \tilde{V}_{j}(\mathbf{r}_{2};\omega)
\vert \phi_{q} \rangle	
=
\sum_{m=0}^{\infty}
\sum_{k=0}^{m}
(-1)^{k}
\dfrac{\lambda_{i}^{2k}}{k!}
\phi_{m,i} (\mathbf{r}_{1})\phi_{m,i} ^{*}(\mathbf{r}_{2}).
\end{eqnarray}
\noindent
By using the photon number operator, which satisfies $a^{\dagger}a\vert m \rangle=m \vert m\rangle$, it follows from Eq. \eqref{lfi} that the cross spectral density operator, say $\hat{\mu}_{ii}$, is given by 

\begin{eqnarray}\label{mu}
\hat{\mu}_{ii}
=
e^{- 
[ 
\hat{a}_{\mathbf{k}i}^{\dagger}\hat{a}_{\mathbf{k}i}
\mathrm{In} (\lambda_{i}) ]^{2}
}
.
\end{eqnarray}
\noindent
Therefore, the expected value of the cross spectral density function is found to be

\begin{eqnarray}\label{tr}
\mathrm{Tr}\{\hat{\rho}\hat{\mu}_{ii}\}
=
e^{- 
(\alpha^{2}_{\mathbf{k}}+\alpha^{4}_{\mathbf{k}})
\mathrm{In} (\lambda_{i})^{2}
}.
\end{eqnarray}
\noindent
The density operator $\hat{\rho}$ is represented as a superposition of coherent states which form an overcomplete basis but they are non orthogonal, the lack of orthogonality of the coherent state representation also holds and is revealed by the fact that the basis of the eigenvalues power expansion is also non-orthogonal.
\noindent
From Eq. \eqref{tr} it can be seen that if $\lambda \rightarrow 1$ the field is completely coherent and consists of only one mode for a given number of photons, the corresponding CSD can therefore be factorized. If $\lambda \rightarrow 0$ the field consists of a superposition of infinite modes weighted by their corresponding eigenvalues and turns out to be incoherent. The larger the number of modes, the smaller number of photons per mode so that the corresponding CSD can no longer be factorized. In the more general and realistic case, the pump field is partially coherent, the expected value is then formed by a superposition of several modes and depending on the strength of the eigenvalues, the number of photons per mode contribute with different amounts to the field coherence. 
\noindent
It is important to note that down-conversion efficiency relies on the phase matching conditions, when these are fulfilled and plane waves are used as pump, the sinc function attains its maximum value. If the pump field is partially coherent this situation changes, Eq. \eqref{tr} shows that the expected value of the cross spectral density operator attains its maximum value for a larger frequency range centred at the phase matching condition. As a consequence maximum efficiency can be attained for partially coherent beams for a given mean frequency of the pump beam.    

\begin{figure}
\begin{center}
\scalebox{0.8}
{\includegraphics{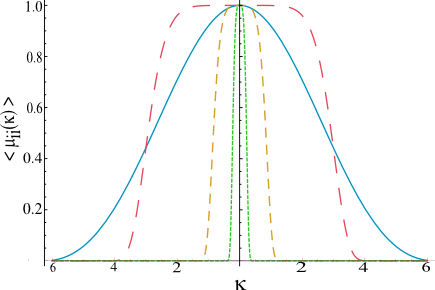}}
\end{center}
\caption{Expected values of the cross spectral density operator for different values of $\lambda$. Cohrent case $\lambda_{i}\rightarrow 1$ is represented by large-dashed curve. Partially coherent case $\lambda_{i}=0.5$ is represented by medium-dashed curve. Incoherent case $\lambda_{i}\rightarrow 0$ is represented by short-dashed curve. For comparison, solid line is the typical sinc function for the coherent case. In figure $\kappa=\mathbf{k}_{0}-\mathbf{k}-\mathbf{k}'$. }
\label{fig1}
\end{figure}

In figure \ref{fig1} we compare the contribution of the typical sinc function versus the expected values of the CSD density operator. It can be seen that for completely coherent fields the sinc function imposes a natural limit for which the states become rather constrained. If partially coherent light is employed, this restriction no longer appears and a number of new states are feasible. In particular for incoherent fields, i.e., when $\lambda_{i}\rightarrow 0$, the two photon amplitude approaches to a delta function, the coherence properties can therefore be engineered a priori opening new possibilities to attain maximally entangled states. In this regard a relationship between entanglement, the two photon amplitude and partially coherent fields will be investigated in the next section.

Now some consequences of the inclusion of the correlation properties will be shown. On substituting Eq. \eqref{tr} into \eqref{g21} the second order correlation function is 

\begin{eqnarray}\label{g21}
\Gamma^{(1)}(\mathbf{r}_{1},\mathbf{r}_{2};\omega)
=
\iint
\delta_{kk'}
\mathrm{Tr}\{\hat{\rho}\hat{\mu}_{ii}\}
F^{*}(\mathbf{k},s,\mathbf{k'},s',\mathbf{r}_{1}t)
F(\mathbf{k},s,\mathbf{k'},s',\mathbf{r}_{2},t)
\mathrm{d}\mathbf{k}
\mathrm{d}\mathbf{k}'.
\end{eqnarray} 
\noindent
The two photon amplitude now exhibits a more involved phenomenon. For $t>\Delta t$ it describes the spectral contribution of coherence to the probability of detecting one photon at $\mathbf{r}$ with frequency $\omega$.

A CSD is called genuine if it satisfies the non negative definiteness condition \cite{9781139644105}, this is a fundamental feature of all physically realizable correlation functions. However not every correlation function associated with an electromagnetic field satisfies this condition. A necessary and sufficient condition states that a correlation function represents a genuine CSD if and only if it can be formally expressed in the form \cite{Gori:07,Gori:09}  	 

\begin{eqnarray}\label{wg}
\mathbf{W}(\mathbf{r}_{1},\mathbf{r}_{2};\omega)
=
\int \int
p(\mathbf{v}_{1},\mathbf{v}_{2})
\mathbf{H}^{*}(\mathbf{r}_{1},\mathbf{v}_{1};\omega)
\mathbf{H}(\mathbf{r}_{2},\mathbf{v}_{2};\omega)
\mathrm{d} \mathbf{v}_{1}
\mathrm{d} \mathbf{v}_{2}.
\end{eqnarray}  

\noindent
This important result in modern coherence theory states that the CSD pertains to a class of functions in a reproducing kernel Hilbert space (RKHS) and satisfies the reproducing property \eqref{fi}. 
\noindent
Since both correlation functions in the integrand of Eq. \eqref{g21} satisfy the non negativity requirement, and because $\langle A_{k}^{\dagger}A_{k'}\rangle=\delta_{kk'}$, their product can be regarded as the product of two reproducing kernels in the more general Hilbert space $H=H_{Q}\otimes H_{S}$. Functions pertaining to the space $H_{Q}$ are functions in the quantum ensemble of states and functions pertaining to the space $H_{S}$ are functions of the statistical ensemble of realizations. By the integral representation theorem for reproducing kernels \cite{opac-b1128469}, the product of the correlation functions in the integrand of Eq. \eqref{g21} is a reproducing kernel. Therefore, on substituting Eq. \eqref{mu} into Eq. \eqref{g21} the function  

\begin{eqnarray}
\Gamma^{(1)}(\mathbf{r}_{1},\mathbf{r}_{2};\omega)
=
\iint
e^{- 
(\alpha_{k}^{2}+\alpha_{k}^{4})
\mathrm{In} (\lambda_{i})^{2}}
F^{*}(\mathbf{k},s,\mathbf{r}_{1},t)
F(\mathbf{k},s,\mathbf{r}_{2},t)
\mathrm{d} \mathbf{k}
\end{eqnarray} 
\noindent
do satisfy the property \eqref{wg} for $\mathbf{v}_{1}=\mathbf{v}_{2}=\mathbf{k}$ and indeed becomes a reproducing kernel. Hence the dual property of the quantum nature of down-conversion process and classical correlations are contained in the new reproducing kernel Hilbert space of the two photon amplitude. The two photon amplitude $\Gamma^{(2)}(\mathbf{r}_{1},\mathbf{r}_{2};\omega)$ contains all the information of the classical and quantum system through the one photon amplitude $\Gamma^{(1)}(\mathbf{r}_{1},\mathbf{r}_{2};\omega)$ which describes the joint contribution of the classical correlation functions to the two photon amplitude of down converted photons. 

It could be tempting a decomposition of the TPA similar to Eq. \eqref{ms}, however, since the TPA in general is not hermitian, a Schmidt decomposition \cite{peres1994} is possible instead. In the next section it will be shown that such a a decomposition implies a link between partially coherent fields and non-classical fields, i.e fields with super Poisson statistics and fields with sub- Poisson statistics.

\section{Partially correlated fields and quantum entanglement}

The results of the previous sections can be exploited to investigate the effect of correlation properties of the pump beam on the two photon wave function.
\noindent
In coherence theory two relationships between the eigenvalues and the properties of the CSD matrix, namely

\begin{eqnarray}
\int \mathrm{Tr}
[
\mathbf{W}^{\dagger}(\mathbf{r}_{1},\mathbf{r}_{2};\omega)
\mathbf{W}^{\dagger}(\mathbf{r}_{1},\mathbf{r}_{2};\omega)
]
\mathrm{d}\mathbf{r}_{1}
\mathrm{d}\mathbf{r}_{2}
=
\sum_{n}\Lambda_{n}^{2},
\end{eqnarray}
\noindent
and

\begin{eqnarray}
\iint \mathrm{Tr}
[
\mathbf{W}^{\dagger}(\mathbf{r},\mathbf{r};\omega)
]
\mathrm{d}\mathbf{r}
=
\sum_{n}\Lambda_{n},
\end{eqnarray}
\noindent
lead to the definition of the effective degree of coherence \cite{1464-4258-6-3-007,1464-4258-9-10-022}

\begin{eqnarray}\label{muef}
\mu_{eff}=
\frac
{\sum_{n}\Lambda_{n}^{2}}
{\big( \sum_{n}\Lambda_{n} \big)^{2}}.
\end{eqnarray}
\noindent
This value represents a relative measure of coherence within a specific area. Note that the term $1/\sum \Lambda_{n}$ corresponds to the Schmidt number which is a measure of entanglement. The effective 	degree of coherence and the Schmidt number seem to play equivalent role for coherence and entanglement.    
In the present case, the effective degree of coherence \eqref{muef} can be calculated from the eigenvalue power expansion \eqref{la}, substituting Eq. \eqref{la} into \eqref{muef} the effective degree of coherence is given by

\begin{eqnarray}
\mu_{eff}=
\frac
{
\sum_{n=0}^{\infty}
\Big(
\sum_{m=0}^{n}
(-1)^{m}
\dfrac{\lambda^{2m}}{2m!}
\Big)^{2}
}
{
\Big(
\sum_{n=0}^{\infty}
\sum_{m=0}^{n}
(-1)^{m}
\dfrac{\lambda^{2m}}{2m!}
\Big)^{2}
}.
\end{eqnarray}
\noindent
In general the eigenvalues depend on the ratio $\lambda=\dfrac{\sigma_{w}}{\sigma_{c}}$ and are determined fundamentally by the analytic form of the CSD matrix. It is clear that for a given genuine CSD this result shows the fact that there exist an effective number of modes with a defined number of photons per mode as stated by Eq. \eqref{tr}.   

\begin{figure}
\begin{center}
\scalebox{1}
{\includegraphics{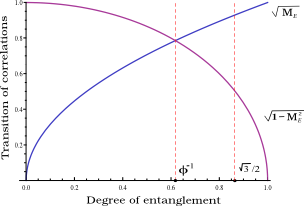}}
\end{center}
\caption{Proportion between the maximal entanglement (blue line) and minimal entanglement (purple line). Region limited by parallel red lines represents the transition zone, where Cauchy-Schwarz inequalities are violated, from super-Poisson statistics for $M_{e}<\phi^{-1}$ to sub-Poisson statistics for $M_{e}>\frac{\sqrt{3}}{2}$. }
\label{figure2}
\end{figure}

It may be tempting to represent the TPA in the form \eqref{ms}, however the lack of hermiticity imposes a rather strong condition for the implementation of a modal expansion therefore the two photon system does not admit a coherent mode representation but a Schmidt decomposition. 
The main distinction between these decompositions relies in the fact that the two photon system Hilbert space comprises the tensor product of Hilbert spaces of functions $\mathbf{u},\mathbf{v}\in H_{Q}$ and $\mathbf{x},\mathbf{y}\in H_{S}$. A pure bipartite system is given by

\begin{eqnarray}
\xi
=
\alpha \mathbf{x}\otimes \mathbf{u}
+
\beta \mathbf{y}\otimes \mathbf{v}.
\end{eqnarray}
\noindent
In the Schmidt basis the density matrix of the system is given by \cite{peres1994}

\begin{eqnarray}\label{sd}
\rho=\xi\xi^{\dagger}
=
\sum_{i,j}c_{i}c^{*}_{j}
\alpha u_{i}u^{\dagger}_{j}\otimes v_{i}v^{\dagger}_{j},
\end{eqnarray} 
\noindent
which can be expressed as the product of reduced density matrices of the form \eqref{ms}. This implies that whereas non hermiticity of operators and non orthogonality of bipartite system basis do not allow us to express Eq. \eqref{1} in the form \eqref{ms}, the reduced density matrix admits a Mercer's series provided hermiticity, orthogonality and square integrability is fulfilled in each subsystem.
\noindent
An expansion of the form \eqref{sd} for entangled states of bipartite systems has been proposed to express the system as the sum of a completely entangled states and a factorizable state \cite{PhysRevA.64.050101}. In the same spirit and by taking into account that for partially coherent fields the Siegert relations holds, we propose a Schmidt decomposition for the TPA as the sum of a factorizable non entangled amplitude function plus a completely entangled amplitude function of the form

\begin{eqnarray}\label{sieg}
\Gamma^{(2)}_{E}(\mathbf{r}_{1},\mathbf{r}_{2})
=
\sqrt{M_{E}}
\Gamma^{1}_{e}(\mathbf{r}_{1},\mathbf{r}_{2})
+
\sqrt{1-M_{E}^{2}}
\Gamma^{1}_{f}(\mathbf{r}_{1},\mathbf{r}_{2}),
\end{eqnarray}
\noindent
with the entangled component 

\begin{eqnarray}\label{4a}
\Gamma^{1}_{e}(\mathbf{r}_{1},\mathbf{r}_{2})
=
\vert
\Gamma^{1}(\mathbf{r}_{1},\mathbf{r}_{2})
\vert ^{2},
\end{eqnarray}
\noindent
and the factorized component

\begin{eqnarray}\label{4b}
\Gamma^{1}_{f}(\mathbf{r}_{1},\mathbf{r}_{2})
=
\Gamma^{1}(\mathbf{r}_{1},\mathbf{r}_{1})
\Gamma^{1}(\mathbf{r}_{2},\mathbf{r}_{2}).
\end{eqnarray}
\noindent
A measure of entanglement that characterizes such a bipartite system is then given by $M_{E}\in \mathbf{C}$ that satisfies $0\leq \vert M_{E} \vert \leq1$.
The corresponding second order correlation functions are related by the Cauchy-Schwarz inequality
\begin{eqnarray}\label{cs}
\sqrt{M_{e}}
\vert
\Gamma^{(1)}(\mathbf{r}_{1},\mathbf{r}_{2})
\vert ^{2}
\leq
\sqrt{1-M_{e}^{2}}
\Gamma^{(1)}(\mathbf{r}_{1},\mathbf{r}_{1})
\Gamma^{(1)}(\mathbf{r}_{2},\mathbf{r}_{2}).
\end{eqnarray}
\noindent
In order that this inequality holds, the inequality $\sqrt{M_{e}}\leq\sqrt{1-M_{e}^{2}}$ must also hold, therefore the degree of entanglement is bounded between the interval 

\begin{eqnarray}
0\leq M_{e}\leq  \dfrac{(-1)\pm\sqrt{5}}{2}.
\end{eqnarray}
\noindent
There are two values for which the inequality holds, one real and one complex, interestingly they are in terms of the \emph{golden ratio} $\phi=\dfrac{1+\sqrt{5}}{2}$.  
\noindent
Since $\phi-1=\frac{1}{\phi} =\dfrac{\sqrt{5}-1}{2} $ in the real case the degree of entanglement satisfies

\begin{eqnarray}
0\leq M_{e}\leq  \dfrac{1}{\phi},
\end{eqnarray}
\noindent
the negative solution, associated with the complex plane, implies that the complex degree of entanglement, denoted by $\tilde {M}_{e}$, satisfies

\begin{eqnarray}
0 \leq \tilde {M}_{e} \leq i \phi.
\end{eqnarray}
\noindent
Beyond this bound, the Cauchy-Schwarz inequality no longer holds, consequently the TPA may undergo changes in its statistics due to entanglement effects. This remarkable result indicates that the degree of entanglement, within its interval of definition, can lead to changes in bi-photon counting statistics. It is interesting not only because there exists an upper bound for entanglement but because establishes a \emph{threshold} between entanglement and the violation of the Cauchy-Schwarz inequality which is a key point for the anti-bunching effect. 
\noindent
The nature of entanglement can be enlightened further by characterising the TPA in terms of the second order correlation functions. On substituting Eqs. \eqref{sieg}, \eqref{4a} and \eqref{4b} into Eq. \eqref{cs} the following pair of inequalities are found

\begin{eqnarray}
2\sqrt{M_{e}}
\vert
\Gamma^{(1)}(\mathbf{r}_{1},\mathbf{r}_{2})
\vert ^{2}
\leq
\Gamma^{(2)}(\mathbf{r}_{1},\mathbf{r}_{2})
\leq 
2\sqrt{1-M_{e}^{2}}
\Gamma^{(1)}(\mathbf{r}_{1},\mathbf{r}_{1})
\Gamma^{(1)}(\mathbf{r}_{2},\mathbf{r}_{2}).
\end{eqnarray}
\noindent
The violation of the Cauchy-Schwarz inequality is a necessary but not sufficient condition for anti-bunching effects hence $M_{e}=\frac{1}{\phi}$ does not guarantee anti-bunching. Specifically the sufficient condition implies that $\Gamma ^{(2)}(\mathbf{r}_{1},\mathbf{r}_{2}) < 1$, beyond this bound the statistics turns out to be sub-Poisson with a pump field originally super-Poisson or Poisson in the coherent case. The value for which this occurs is found to be $M_{e}>\sqrt{3}/2$. Figure \ref{figure2} illustrates the \emph{golden section} $\phi\leq M_{e}\leq \sqrt{3}/2$ where this transition phenomena arise.

This result states that, provided entanglement and coherence properties are properly engineered, anti-bunching effects with partially coherent light are feasible since the fourth order correlation function admits a Schmidt decomposition that can be carried out by means of the coherent-mode representation of the second order correlation functions. It transpires that coherence properties of electromagnetic fields can be used for tailoring the two photon amplitude for quantum entanglement strategies and vice versa, quantum entanglement can be used to manipulate the photon counting statistics of partially coherent down-converted fields. In addition, the obtained result suggests that the non-genuine correlation functions, those for which the non negativity condition \eqref{w3} is not satisfied, can be studied in connection with non-classical fields, i.e., those for which anti-bunching effects occur, provided the appropriate statistics for photon counting events is selected.
The presented results indicate that correlation properties of the pump field introduces a new degree of freedom for tailoring the two photon amplitude in entanglement applications, end moreover entanglement can be engineered to test transition states between super-Poisson to sub-Poisson statistics.

\section{Conclusions}

We have shown that the coherence properties of the electromagnetic pump field are preserved under parametric down-conversion. It was demonstrated that the two photon amplitude contains full information about both quantum down-converted photons and the classical correlation properties of the pump beam. By using the coherent mode representation, the statistical nature of the CSD matrix has been transferred to a CSD operator whose expectation value allows us to determine the shape of the TPA for a partially coherent pump field. The relationship between the expected value and the kernel properties of the CSD led to the deduction of the reproducing kernel property of the second order correlation function. This property and the modal decomposition account for a complete characterization of the TPA in terms of coherent modes. As a consequence, by choosing appropriately an orthonormal basis of the pump beam, the TPA can be shaped in terms of the number of modes resulting in a new strategy for entanglement engineering. 
We put forward a measure of entanglement by means of the TPA as a direct extension of the entanglement measure based on states \cite{PhysRevA.64.050101}. This measure allows us to establish upper bounds for the degree of entanglement, the most important conclusion is that field correlations through the modal decomposition can be deployed to tailoring the TPA and conversely the degree of entanglement can be used to test the transition of fields obeying super-Poisson statistics to sub-Poisson, therefore anti-bunching effects are expected provided the appropriate correlation properties are engineered. The presented results indicates that correlation properties of the pump field introduce a new degree of freedom for down-converted fields. This results can be useful in quantum communications and quantum information theory.






\section{Acnowledgements}

acknowledgments The authors acknowledge Sir Peter Knight, Stephen Barnett and Neal Radwell for valuable comments and suggestions. M. A. Olvera gratefully acknowledges CONACYT for financial support under grant 232259.

\bibliographystyle{ieeetr} 

\bibliography{MiguelTPA}

\end{document}